\documentclass[a4paper]{article}
\pdfoutput=1
\usepackage{WCSB,amsmath,amssymb, epsfig}
\usepackage{comment}

\title{SPATIAL ORGANIZATION OF PROTEOMES: A LOW-RANK APPROXIMATION}
\name{Federico Felizzi$^{1}$,  Jerome Galtier$^2$,  Georgios Fengos$^{1}$, Dagmar Iber$^{1}$ }
\address{$ ^1$D-BSSE, ETH Zurich, Mattenstrasse 26, 4058 Basel, Switzerland, \\
$ ^2$France Telecom, 905 rue Albert Einstein, 06921 Sophia Antipolis Cedex, France\\ 
federico.felizzi@bsse.ethz.ch, jerome.galtier@orange-ftgroup.com, \\georgios.fengos@bsse.ethz.ch, dagmar.iber@bsse.ethz.ch}
\begin{document}

\maketitle
\begin{abstract}
We investigate the problem of signal transduction via a descriptive analysis of the spatial organization of the complement of proteins exerting a certain function within a cellular compartment. \\
We propose a scheme to assign a numerical value to individual proteins in a protein interaction network by means of a simple optimization algorithm. We test our procedure against datasets focusing on the proteomes in the neurite and soma compartments. 
\end{abstract}

\section{INTRODUCTION}
\label{sec:intro}
In response to external stimuli, cells undergo a series of biochemical reactions. Proteins are the key players in such processes. Protein signalling networks are the entities responsible for transmitting the signal from external cues. Advancements in experimental methods, along with the availability of continuous improvements of protein-protein interaction resources, enable the construction of experimentally driven networks. Protein networks typically harbour few highly connected nodes, whose knockout alters the overall function of the biological network. A number of approaches to integrate (phospho)proteomics data and protein-protein interaction databases have been proposed \cite{Chuang:2007p16024, Klammer:2010p14882}.  Results from phosphoproteomics experiments are mapped onto the STRING database \cite{Jensen:2009p9151} to identify differentially regulated subnetworks \cite{Klammer:2010p14882}.  Understanding the global effects of protein inhibition leads to the prediction of the change of a cell's behaviour in response to perturbations of the signaling network. Although post-translational protein modifications are often depicted to follow a specific time ordering \cite{Cobb:1999p16204}, most large-scale experimental analyses fail to deliver a large number of time samples, calling for a methodology that aims at understanding the interactome as a whole.  Graph theoretical procedures have been used extensively in understanding the importance of protein and protein complexes from the topology of a protein interaction network \cite{Aittokallio:2006p14742}. Centrality measures such as \textit{degree centrality} play a very important role in understanding the importance of a node in the entire network \cite{Estrada:2006p14730}. It was argued that centrality measures are important in identifying nodes as potential drug targets. Nonetheless, protein-protein interactions present highly connected nodes, whose inhibition results in failure of the entire network \cite{Albert:2000p14668}. Our interest here goes to the development of an algorithm to understand the global structural property of a protein interaction network resulting from proteomics experiments. In this work our focus is the spatial organization of proteins during neurite extension \cite{Pertz:2008p12973}. \\
We used proteomics measurements  that identified 4855 proteins from the soma and neurite proteomes of neuroblastoma cells \cite{Pertz:2008p12973}. As discussed in \cite{DaSilva:2002p14352}, Cdc42 and Rac1 have been identified as the major player in the organization of the acting cytoskeleton in response of extracellular cues. The bioinformatics analysis performed revealed the complement of proteins that intermediate Rac1 and Cdc42 signaling. 
Counting the relative abundance of a given protein sequence in independent mass spectrometry measurements for the soma and the neurite protrusion, revealed the relative abundance of a given protein in each of the samples. Such approach identified 1229 proteins that were enriched in the neurite. Out of those, about 800 proteins revealed an enrichment value larger that $2.0$. Of those 800, about 200 were mapped onto a network via the IPA resource (IPA), and 36 were mapped on a potential interactome via the Babelomics resource \cite{Medina:2010p16118}. 
In this work we combine experimental data with knowledge from protein-protein interaction databases to construct a weighted graph. The weight on an edge is seen as the strength of the interaction between two proteins. Point on the location of a manifold represent proteins. We construct a quadratic potential, in that point on the surface repel each other according to the value of the weight connecting them. All the positions of the points are progressively updated, until the variation of the potential reaches a plateau. We show how the equilibrium configuration reveals insight on the role of the studied proteins in the context analysed. Furthermore, the analysis of the most relevant subspaces suggests constraining the problem to $S^2$, the 2-dimensional sphere.

\section{METHODS}
\label{sec:general}
IPIs (International Protein Indexes) and the relative level of enrichment of the neurite-enriched proteins are extracted from published data \cite{Pertz:2008p12973}. A weighted graph $ G = (V,E)$  is constructed by using information from the STRING database \cite{Jensen:2009p9151}. 
\subsection{Network Construction}
The STRING database \cite{Jensen:2009p9151} presents comprehensive information on the likelihood of a binary protein-protein interaction for various organisms. The reliability of an interaction is presented as a number $w_{\mbox{\tiny{DB}}}(e) \in [0,1]$, where $e = (p_i, p_j)$ is the edge connecting two protein nodes. We filter data from Proteomics experiments (enrichment/depletion) with the edge weights suggested by the STRING database \cite{Jensen:2009p9151} to build an undirected weighted graph object. The rationale in building the graph is to assign a weight on an edge connecting two proteins - as observed in the database - in the following fashion:
\begin{equation}
\label{eqn:weights}
w(p_1, p_2) =  w_{\mbox{\tiny{DB}}}(e) \cdot v^*(p_1)  \cdot v^* (p_2),   
\end{equation}
assuming causality in simultaneous up/down regulation, 
\begin{equation}
v^*(p_1) = \left\{ \begin{array}{ll} v(p_1) & \mbox{if}~ v(p_1) \geq 1 \\ \frac{1}{v(p_1)} & \mbox{if}~ v(p_1) \leq 1 \end{array} \right.
\end{equation}
where $v(p_i)$ are the levels of enrichment of protein $p_i$ reported in the supplementary material of \cite{Pertz:2008p12973}. 
\subsection{Problem Formulation}
\label{sec:probform}
Let $X_i$ be the vector of the coordinates of point $i$ in the $d$-dimensional Euclidean space $\mathbb{R}^d$. Let $w_{ij} \in \mathbb{R}$ be the score on the interaction between protein $p_i$ and protein $p_j$ as defined by $w(p_i,p_j)$ in equation \ref{eqn:weights}. We aim at identifying the optimal location of the points $X_i$ by solving the optimization problem \ref{eqn:ProblemFormulation}. 
\label{ssec:titlestyle}
\begin{equation}
\label{eqn:ProblemFormulation}
\begin{array}{rl}
\mbox{max}&\sum_{ij} w_{ij}\parallel X_i - X_j\parallel^2\\
\mbox{s.t.}&{\parallel X_i \parallel}^2 = 1, \forall i=1,\ldots,N\\
~&X_i \in \mathbb{R}^d
\end{array}
\end{equation}
Note that different constraints on the values values of $d$ correspond to different rank constraints for the matrix $\mathbf{X} \in \mathbb{R}^{N \times d}$ representing the coordinates of the points. The case $d = N$ and $d = 0$ are the two extreme case, the latter corresponding to assign only binary labels $\{-1,+1\}$. In section \ref{sec:rank}, we discuss on the reasons that led us to sett $d = 3$ as rank constraint for any subsequent operation we performed on the matrix $\mathbf{X}$. 
\subsection{Structure of the Algorithm}
Proteins are seeded to locations on the surface of the $(d-1)$-dimensional sphere $S^{d-1}$. A loop over the points is performed. The positions of each of those is updated according to \ref{eqn:Algo}. 
\begin{center}
\begin{minipage}[htbp]{10cm}
\label{eqn:Algo}
\begin{tabbing}
Start with a feasible point $X$.\\
\textbf{while} ~~ $\|f(X^{\mbox{old}})-f(X^{\mbox{new}})\|>\epsilon$\\
~~~~\textbf{for}~~i = $1$~to~$N$\\
~~~~~~~~$X_i^{\mbox{new}}=-\frac{\sum_j w_{ij}X_j^{\mbox{\tiny{old}}}}{|\sum_j w_{ij}X_j^{\mbox{\tiny{old}}}|}$\\
~~~~\textbf{end}\\
\textbf{end}
\end{tabbing}
\end{minipage}
\end{center}
here $f(\mathbf{X}) = \sum_{ij} w_{ij}\parallel X_i - X_j\parallel^2$ and $\epsilon$ is an arbitrary value (we set it to 0.01) that acts as a threshold on the variation of the underlying potential. 
Upon updating the value of $X_i^{\mbox{new}}$, the minus sign results from the KKT conditions \cite{ConvOPT}. The optimality condition is
\begin{equation}
\sum_j w_{ij}(X_i-X_j) + \lambda_i X_i = 0,~~~\|X_i\|^2=1, 
\end{equation}
giving
\begin{equation}
X_i=\pm \frac{\sum_j w_{ij}X_j}{|\sum_j w_{ij}X_j|},
\end{equation}
where $\lambda_i$ are the Lagrange multipliers and the minus sign comes from the sign of the Hessian matrix. 
The minimum formulation of the optimization problem \ref{eqn:ProblemFormulation} converges to a global  minimum up to symmetries \cite{KatzCooper} . 
The application of the algorithm \ref{eqn:Algo} results in points reaching an equilibrium on the surface of a sphere.\\
Removal of one link, or of one point and all its incident edges will result in a new equilibrium for the system. 
\begin{figure}[t]
\centering
\includegraphics[width=7cm, height=4.5cm]{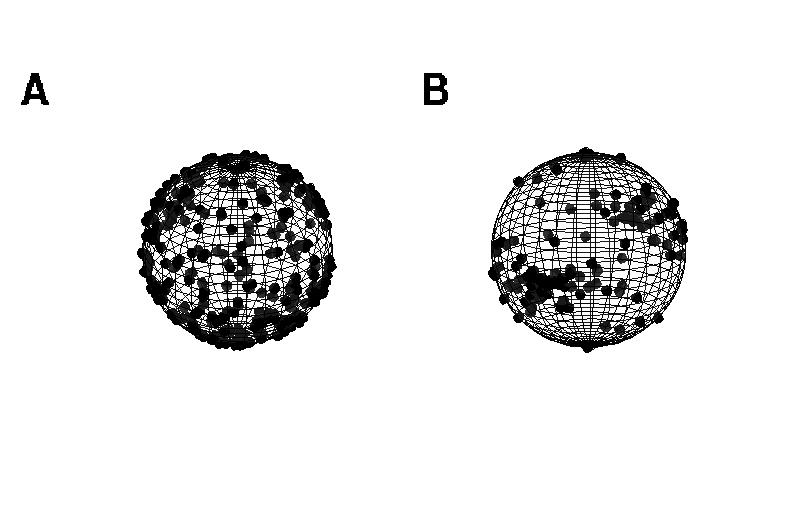}
\caption{Initial (A) and Final (B) configurations of the points on the surface of the manifold. Here $S^2$ is chosen for visualization. In (A) the points are distributed uniformly, in (B) they occupy the equilibrium positions following the application of the algorithm \ref{eqn:Algo}. }
\label{fig:INI_FIN}
\end{figure}
\subsection{Euclidean Distance Matrix (EDM)}
\label{sec:EDM}
\begin{figure}[t]
\centering
\includegraphics[width=7cm, height=4.5cm]{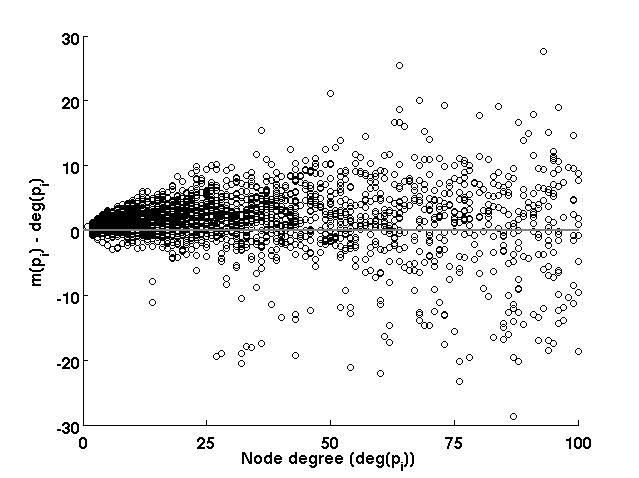}
\caption{Correlation between the score defined by equation \ref{eqn:funscore} and the degree of the node. Large positive values of $m(p_i) - deg(p_i)$ exhibit potential knockout targets}
\label{fig:relevance}
\end{figure}

Starting from the final configuration of points at equilibrium $\mathbf{X}^{\mbox{\tiny{final}}}$, we construct the Euclidean Distance Matrix $E$ (\cite{Dattorro:2005p14477}) as
\begin{equation}
E_{ij} = || X_i^{\mbox{\tiny{final}}} - X_j^{\mbox{\tiny{final}}}||_2 
\end{equation}
where $X_i^{\mbox{\tiny{final}}}$ and $X_j^{\mbox{\tiny{final}}}$ are the $d-$dimensional vectors corresponding to the equilibrium locations of points $i$ and $j$. 
Let $E$ be the Euclidean Distance Matrix and let $S$ be the symmetric adjacency matrix of the protein interaction graph $G = (V,E)$ defined as
\begin{equation}
S_{ij} =\left\{ \begin{array}{ll} 1 & \mbox{if}~ (i,j) \in E \\ 0 & \mbox{otherwise} \end{array} \right.
\end{equation}
 Let us introduce the matrix
\begin{equation}
H = S \circ E
\end{equation}
where $\circ$ denotes the Hadamard product. Let $h_i$ be the $i$-th column of matrix $H$. We define a score on protein $i$ as 
\begin{equation}
\label{eqn:funscore}
m(p_i) = \mathbf{1}^T h_i
\end{equation}
The score defined by equation \ref{eqn:funscore} provides a measure on the relevance of protein $p_i$ in the specific experimental context. We analyzed the values $m(p_i)$ for the entire neurite proteome \cite{Pertz:2008p12973} . We looked at the correlation between the degree of a protein in the network $deg(p_i)$ and its associated score $m(p_i)$. The results of the analysis restricted to low-degree proteins is shown in figure \ref{fig:relevance}. We argue that high positive drifts of $m(p_i)$ from $deg(p_i)$ are likely to identify proteins having a high relevance in a specific biological context. \\
Removal of one point perturbs the equilibrium and relaxes the system to a new optimal configuration. We analyzed the effects of removing each seeded point with respect to Rac1 and Cdc42. For this purpose, we introduced the matrix $\mathbf{\tilde{X}}^{\mbox{\tiny{init}}} \in \mathbf{R}^{(N-1) \times d}$, such that $\mathbf{\tilde{X}}^{\mbox{\tiny{init}}} = \mathbf{{X_{[-i]}}}^{\mbox{\tiny{final}}}$. Here $\mathbf{{X_{[-i]}}}^{\mbox{\tiny{final}}}$ denotes the removal of the $i$-th row from matrix $\mathbf{{X}}^{\mbox{\tiny{final}}}$. Let ${v_0^j}^T$ and ${v_{[-i]}^j}^T$ be the final equilibrium location of points $j$ (representing either Cdc42 or Rac) before and after removing point $i$. Let us introduce the matrices
\begin{equation}
D^j = \left[ \mathbf{{X_{[-i]}}}^{\mbox{\tiny{final}}} - \left[ \begin{array}{c} {v_0^j}^T \\ \vdots\\  {v_0^j}^T \\ \end{array} \right] \right] \in \mathbf{R}^{(N-1) \times 3}
\end{equation}
and
\begin{equation}
D_{[-i]}^j = \left[ \mathbf{{\tilde{X}}}^{\mbox{\tiny{final}}} - \left[ \begin{array}{c} {v_{[-i]}^j}^T \\ \vdots\\  {v_{[-i]}^j}^T \\ \end{array} \right] \right] \in \mathbf{R}^{(N-1) \times 3}
\end{equation}
We introduce a functional score for protein $p_i$ with respect to protein $p_j$ as
\begin{equation}
\label{eqn:fun_score}
fs_j(i) = \mbox{Tr} \left( \left[D^j - D_{[-i]}^j\right]^T \left[D^j - D_{[-i]}^j\right]\right)
\end{equation}
\section{EXPERIMENTAL DATA} 

\subsection{Experimental-driven Network}
We followed the discussion in \cite{Pertz:2008p12973} and isolated the neurite proteome and the 36 GEFs, GAPs and Effectors involved in the Rac1 and Cdc42 interactome. We used weights on the edges given from database and experimental evidence as proposed in the STRING database \cite{Jensen:2009p9151}. 
\subsection{Work related knockdowns}
The work discussed in \cite{Pertz:2008p12973} performed knockdowns and evaluated changes in the behaviour of the neurite formation. The knockdowns performed involved 10 GEFs and GAPs. In table 1 we list the 6 out of those 10 we could map to the unique connected graph component. Changes in the neurite dynamics were observed upon knocking down each individual protein. We list a summary of the experimental observation and the results of our methodology in table 1.

\section{RESULTS}
\subsection{Comparison of different rank constraints}
\label{sec:rank}
The motivation to come up with a low-rank approximation - i.e. the constraint $d = 3$ as discussed at the end of section \ref{sec:probform} - of the spatial organization of the proteomes was related to the importance of the subspaces in the final configuration of the points after the application the algorithm. Let $\mathbf{X}^{\mbox{\tiny{init}}} \in \mathbb{R}^{N \times d}$ and $\mathbf{X}^{\mbox{\tiny{final}}} \in \mathbb{R}^{N \times d}$ be the configuration of the points on the surface of a $d$-dimensional manifold at the beginning and at the end of the simulation respectively. 
Let 
\begin{equation}
\mathbf{X}^{\mbox{\tiny{init/final}}} = U_{\mbox{\tiny{init/final}}} \Sigma^{\mbox{\tiny{init/final}}} V_{\mbox{\tiny{init/final}}}^T
\end{equation}
be the singular value decompositions of the initial and final configuration of points.

 We started simulation with different rank constraints, ranging from $d = N$, whenever computationally feasible, down to $d = 0$. The analysis of the singular values for the results of the optimal configuration of the points after the application of the algorith \ref{eqn:Algo} revealed that 
 \begin{equation}
\sigma_1 > \sigma_2 > \sigma_3 \gg \sigma_4 > \ldots > \sigma_d
 \end{equation}
 being $\sigma_i$ the entries of the matrix $\Sigma^{\mbox{\tiny{final}}}$. Cutting off the dimensionality to $d = 3$ did not affect the dependency of the results with respect to the initial conditions and led to a significant time saving with respect to values $d > 10$. We reason that dimensionality constrains of $d = 1$ or $d = 2$ might lead to an additional time saving when compared to $d = 3$, but the dependency on the initial conditions will be quite significant. If points are placed on a line or a circle, they will occupy specific mutual positions, such as point $p_i$ \textit{is to the right of point} $p_j$ and the nature of the algorithm \ref{eqn:Algo} prevents points from being overlapped, thus switching their locations. More specifically, in $d = 1$ or $d = 2$, if point $p_i$ is adjacent to point $p_j$and $p_i$ \textit{is to the right of} $p_j$, their mutual position will remain such in the entire course of the simulation. A rank constraints of $d = 3$ allows for more spatial freedom, since points do not have to overlap in order to change their mutual positions.

\subsection{Analysis of the EDM - Measure of Relevance}
\label{sec:analEDM}

We summarize the results obtained for the 6 proteins knocked down in the experiments outlined in \cite{Pertz:2008p12973} in table 1. According to the notation of the supplementary materials of \cite{Pertz:2008p12973}, we refer to the GTPase specificity of the listed GEFs and GAPs. The signs in the \textit{Dynamics} column correspond to the observed neurite dynamics. 0 indicates that no neurite dynamics is observed. The sign `+' indicates increased persistence with some protrusion/retraction events. The sign `++' denotes increased persistence with total loss of protrusion/retraction events. For \textit{Trio} an unstable neurite dynamics was observed. The numerical values for the functional scores as defined in equation \ref{eqn:fun_score} exhibit a significant correlation to the observed neurite dynamics. 

\begin{small}
\begin{table}[t]
\label{table:knockdowns}
\centering
\begin{tabular}{|c|c|c|c|}
\hline
\hline
\bf{\small{Name}} & \bf{\small{Dynamics}} & \bf{\small{Spec.}} &\bf{\small{Value} $fs_j(i)$} \\
\hline
Arhgap17 & + & both & 7.61\\
\hline
Arhgap21 & 0 & Cdc42 & 2.98 \\
\hline
ITSN1 & ++ & Cdc42 & 10.29 \\
\hline
Srgap2 & + & Cdc42 & 4.22 \\
\hline
Trio & 0/unstable & Rac1 & 3.87 \\
\hline
Vav3 & 0 & both & 1.00 \\
\hline
\hline
\end{tabular}
\caption{Comparison of the variation in neurite dynamics with the measure on the specific protein introduced in the manuscript. Numerical values are strongly associated with the change in neurite dynamics upon knocking down the annotated proteins. `0', `+' and `++' indicate the observed neurite dynamics as described in section \ref{sec:analEDM}}
\end{table}
\end{small}
\section{CONCLUSION}
Our approach proposes a way to use the integration of different information sources into a unique computational framework. We use the relative levels of enrichment or depletion of protein levels from LC-MS/MS \cite{Pertz:2008p12973} and mapped onto the reliability of a binary protein-protein interaction given by the STRING database \cite{Jensen:2009p9151}. \\
In this fashion, we were able to assign weights on the edges defining protein-protein interactions in an experimentally driven fashion. Inspired by the work performed in \cite{KatzCooper}, we proposed a geometrical interpretation of the spatial organization of a proteome. Such formulation enables the construction of specific measures for each individual protein in the network as a whole. It constitutes a step towards a system oriented analysis, complementing insights on the biological information given by the topology of interaction networks. The numerical values we obtained were well in line with the experimental observations discussed in \cite{Pertz:2008p12973}. Furthermore, the comparison of our score to the degree of a protein in the network might constitute a relevant measures in the identification of drug targets. \\
Possible extensions of this work might include the integration of different database sources, e.g. \cite{Turner:2010p14439}, containing information about protein complexes.

\section{ACKNOWLEDGMENTS}
We thank Philipp Germann and Simon Tanaka for fruitful discussion and reviewing the manuscript. 

\bibliographystyle{IEEEbib}
\bibliography{Prots_Low_Rank}

\begin{thebibliography}{10}

\bibitem{Chuang:2007p16024}
H.~Chuang and et. al.,
\newblock ``Network-based classification of breast cancer metastasis,''
\newblock {\em Molecular Systems Biology}, vol. 3, 2007.

\bibitem{Klammer:2010p14882}
M.~Klammer and et. al.,
\newblock ``Identifying differentially regulated subnetworks from
  phosphoproteomic data,''
\newblock {\em BMC BIOINFORMATICS}, vol. 11, pp. 351, 2010.

\bibitem{Jensen:2009p9151}
L.~Jensen and et. al,
\newblock ``{STRING 8--a global view on proteins and their functional
  interactions in 630 organisms},''
\newblock {\em Nucleic Acids Research}, 2009.

\bibitem{Cobb:1999p16204}
M.~Cobb,
\newblock ``{MAP kinase pathways},''
\newblock {\em Progress in biophysics and molecular biology}, vol. 71, 1999.

\bibitem{Aittokallio:2006p14742}
T.~Aittokallio and B.~Schwikowski,
\newblock ``Graph-based methods for analysing networks in cell biology,''
\newblock {\em BRIEFINGS IN BIOINFORMATICS}, vol. 7, pp. 243--255, 2006.

\bibitem{Estrada:2006p14730}
E.~Estrada,
\newblock ``Virtual identification of essential proteins within the protein
  interaction network of yeast,''
\newblock {\em Proteomics}, vol. 1, pp. 35--40, 2006.

\bibitem{Albert:2000p14668}
J.~H. Albert, R. and A.-L. Barabasi,
\newblock ``Error and attack tolerance of complex networks,''
\newblock {\em Nature}, vol. 406, pp. 378--382, 2010.

\bibitem{Pertz:2008p12973}
O.~Pertz and et~al.,
\newblock ``{Spatial mapping of the neurite and soma proteomes reveals a
  functional Cdc42/Rac regulatory network.},''
\newblock {\em PNAS}, vol. 3, pp. 1931–1936, 2008.

\bibitem{DaSilva:2002p14352}
J.~S. Da~Silva and C.~G. Dotti,
\newblock ``Breaking the neuronal sphere: Regulation of the actin cytoskeleton
  in neuritogenesis,''
\newblock {\em Nature Reviews Neuroscience}, vol. 3, pp. 694--704, 2002.

\bibitem{Medina:2010p16118}
I.~Medina and et. al.,
\newblock ``Babelomics: an integrative platform for the analysis of
  transcriptomics, proteomics and genomic data with advanced functional
  profiling,''
\newblock {\em Nucleic Acids Research}, vol. 38, 2010.

\bibitem{ConvOPT}
S.~Boyd and L.~Vandenberghe,
\newblock {\em Convex Optimization},
\newblock Cambridge University Press, 2004.

\bibitem{KatzCooper}
I.~N. Katz and L.~Cooper,
\newblock ``Optimal location on a sphere,''
\newblock {\em Comp. Maths. with Appls.}, vol. 6, pp. 175--196, 1980.

\bibitem{Dattorro:2005p14477}
J.~Dattorro,
\newblock {\em Convex Optimization and Euclidean Distance Geometry},
\newblock Meboo Publishing, 2005.

\bibitem{Turner:2010p14439}
B.~Turner and et~al.,
\newblock ``irefweb: interactive analysis of consolidated protein interaction
  data and their supporting evidence.,''
\newblock {\em Database: The Journal of Biological Databases and Curation},
  2010.

\end{thebibliography}

\end{document}